# Enhancing Light Harvesting by Hierarchical Functionally Graded Transparent Conducting Al-doped ZnO Nano- and Mesoarchitectures

*Paolo Gondoni[1], Piero Mazzolini[1,2], Valeria Russo[1], Annamaria Petrozza[2], Avanish K. Srivastava[3], Andrea Li Bassi[1,2], and Carlo S. Casari[1,2] \**

[1]Dipartimento di Energia and NEMAS – Center for NanoEngineered Materials and Surfaces, Politecnico di Milano
Via Ponzio 34/3, 20133 Milano, Italy
E-mail: carlo.casari@polimi.it

[2]Center for Nano Science and Technology @Polimi, Istituto Italiano di Tecnologia
Via Pascoli 70/3, 20133 Milano, Italy

[3]CSIR - National Physical Laboratory
New Delhi, 110012, India

ABSTRACT

A functionally graded Al-doped ZnO structure is presented which combines conductivity, visible transparency and light scattering with mechanical flexibility. The nano and meso-architecture, constituted by a hierarchical, large surface area, mesoporous tree-like structure evolving in a compact layer, is synthesized at room temperature and is fully compatible with plastic substrates.

Light trapping capability is demonstrated by showing up to 100% improvement of light absorption of a low bandgap polymer employed as the active layer.

## 1. Introduction

Transparent Conducting Oxides (TCOs) are widely employed in a variety of fields including photovoltaic devices.[1–3] The attention of research has recently been drawn by environmental and economic reasons towards indium free TCOs, such as Al-doped ZnO (AZO).[4–6]

Besides conventional TCO applications which require high transparency and conductivity,[7] the advent of organic and hybrid solar cells has recently caused the rise of interest in developing peculiar functional properties of photoanodes, such as large effective surface area,[8] and efficient light management.[9] To obtain such properties, several strategies have been explored, like chemical etching,[10,11] roughness control,[12] nanowire layers,[13,14] nanotexturing induced by crystal orientation or by direct laser interference patterning,[15,16] plasmonics,[9,17] partial embedding of TCO nanoparticles.[18]

However, these techniques to induce light trapping include either deposition processes or post-treatments at high temperature, or even chemical treatments unsuitable for organic materials. The design of a material by a bottom-up approach allowing to control the individual building blocks and their assembly can provide an alternative solution to this problem; one versatile technique to achieve this is constituted by Pulsed Laser Deposition (PLD), as demonstrated for other oxides.[19–21] Seeking compatibility with organic cells demands that the depositions be performed at room temperature: few reports of such process can be found in literature.[22–26]

We have recently reported the synthesis of compact transparent conductive AZO by room temperature PLD,[26] and the growth of mesoporous AZO films with significantly high light scattering properties.[27,28] In those works we investigated the effects of the oxygen pressure in the deposition chamber on the AZO films, and demonstrated successful adhesion on polymer substrates.[29]

In this work we present a new approach to combine electrical conductivity, optical transparency and haze into one structure suitable for organic photovoltaics. The basic idea is to obtain a functionally graded film, constituted by a mesoporous transparent scattering layer evolving in a compact TCO layer, during the same deposition process, which is fully compatible with flexible substrates. We will show how this approach allows to obtain a sheet resistance of the order of $100 \ \Omega \ \square^{-1}$, visible transparency of the order of 80-90% and haze of 40-50%.

## 2. Material and Methods

AZO was deposited by Pulsed Laser Ablation of a solid $Al_2O_3(2\%wt.):ZnO$ target by a ns-pulsed laser ($\lambda = 266$ nm). The background $O_2$ pressure was varied from 2 to 160 Pa as discussed in the main text, with total deposition times ranging from 30 minutes to 90 minutes to adjust the relative thickness of the porous and compact layers. Depositions were performed on (100) silicon, glass and plastic (Ethylene-Tetrafluoroethylene, ETFE) substrates. The laser fluence on the target was about 1 J cm$^{-2}$ and the target to substrate distance was 50 mm.

PCPDTBT polymer was obtained by Konarka Technologies (molecular mass 27700 g/mol; polydispersity index 1.9; purity >95%). Solutions of PCPDTBT were prepared with a concentration of 10 mg/mL of PCPDTBT in chloroform. The solutions were deposited by spin coating at 1000rpm for 45 seconds in a dynamic mode.

Electrical properties were measured in the 4-point probe configuration with a Keithley K2400 Source/Measure Unit as a current generator (from 100 nA to 10 mA), an Agilent 34970A voltage meter and a 0.57 T Ecopia permanent magnet.

Optical transmittance, haze and absorption spectra were acquired with a UV–vis–NIR PerkinElmer Lambda 1050 spectrophotometer with a 150 mm diameter integrating sphere. Scanning Electron Microscope Images were acquired with a Zeiss SUPRA 40 field emission SEM. High resolution transmission electron microscopy was performed by employing A FEI Tecnai G2 F30 STWIN with field emission gun, operated at the electron accelerating voltage of 300 kV. AZO samples were transferred on carbon coated copper grids for TEM analysis.

## 3. Results and discussion

### 3.1 Synthesis of compact and porous AZO layers

Examples of cross-sectional scanning electron microscope images of AZO films are reported in **Figure 1**. The images show how film morphology can be tuned from compact films with a columnar structure (2 Pa $O_2$, figure 1.a) to porous, hierarchically assembled morphologies (160 Pa $O_2$, figure 1.c), with the possibility to obtain intermediate nanoporous structures (100 Pa $O_2$, 1.b). As discussed in those works, low (0.01 Pa - 10 Pa) $O_2$ deposition pressures yield atom-by-atom deposition of compact, oxygen-deficient films whose resistivity is minimized ($4.5 \times 10^{-4} \Omega$ cm, sheet resistance $\approx 9 \; \Omega \; \square^{-1}$) at 2 Pa $O_2$, where dopants (intentional and unintentional) and oxygen vacancies favor the filling of the conduction band with free electrons, but defect concentration is sufficiently low not to decrease carrier mobility. X-Ray Diffraction showed a nanocrystalline structure with mean domain size of the order of 30 nm, with a strongly preferential orientation along the 0002 planes. In **figure 2**.a, a Transmission Electron Microscopy (TEM) image reveals an atomic scale image of an elongated-columnar morphology in the form of a wire with a diameter of about 9 to 10 nm. A set of atomic planes of hexagonal-ZnO along 0002 are marked with interplanar spacing of 0.26 nm (Fig. 2.a). A corresponding Fast Fourier Transform (FFT) calculated from the lattice scale image also reveals the presence of prominent (0002) planes in the reciprocal space (inset (A)). Inset (B) shows a well resolved interface between the nanowire and the amorphous structure at lattice scale. Inset (C) shows a low magnification image, which elucidates a cluster of elongated morphologies dispersed in the amorphous phase.

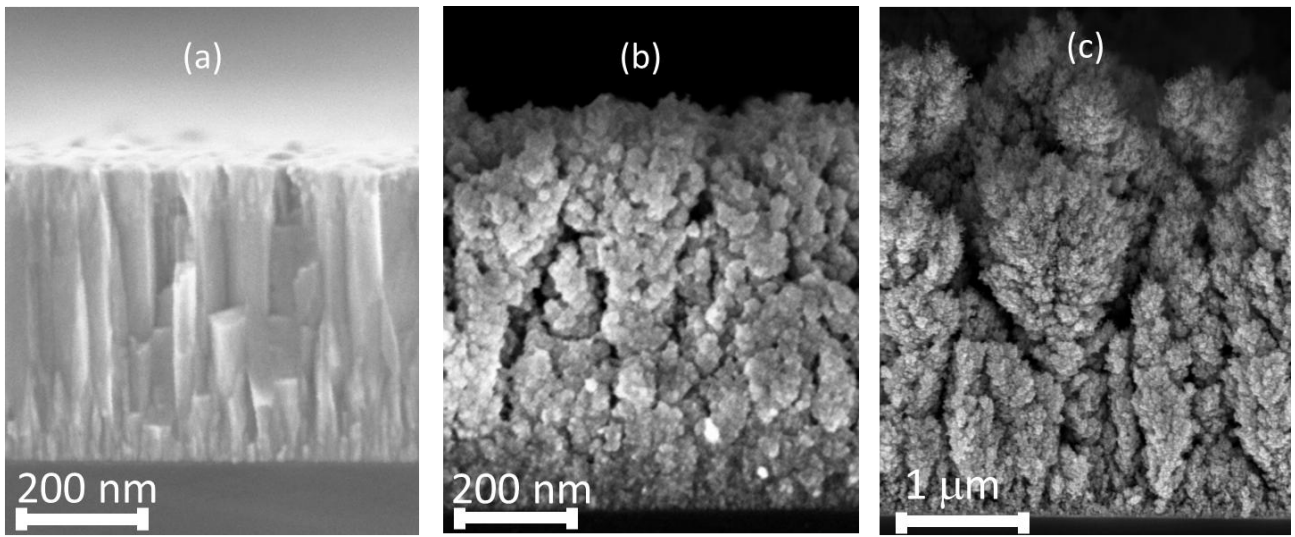

**Figure 1.** Cross-sectional SEM images of a compact sample grown at 2 Pa (a), a nanoporous

sample grown at 100 Pa (b) and a mesoporous sample grown at 160 Pa (c).

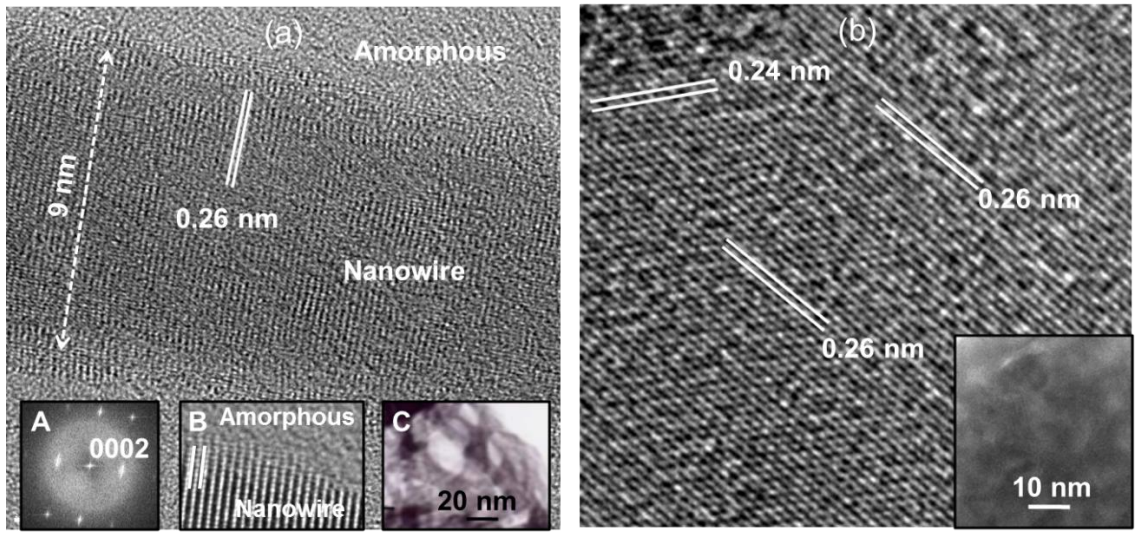

**Figure 2.** (a) Main image: high resolution TEM image of a columnar film grown at 2 Pa, showing preferentially oriented wires surrounded by an amorphous region. Inset A: FFT of the main image, highlighting the (0002) plane position in the reciprocal space. Inset B: Interface between the columnar and amorphous regions. Inset C: Low resolution TEM image showing a cluster of elongated growth within an amorphous region.

(b) High resolution TEM image of a mesoporous film grown at 100 Pa, showing randomly oriented domains. The inset shows an average domain size of about 10 nm.

Higher (100 Pa - 200 Pa) O₂ deposition pressures induce clustering phenomena in the ablation plume, which result in the deposition of mesoporous hierarchical assemblies of nanosized clusters. XRD (not shown) and TEM analyses revealed that the nanostructure of the films is constituted by nanocrystalline domains of about 10 nm in size (as shown in figure 2.b) with random orientation. Sets of atomic planes along $0002$ and $10\bar{1}0$ with interplanar spacings of 0.26 nm and 0.24 nm, respectively, are marked on the micrograph (Fig. 2.b). The inset in figure 2.b exhibits the distribution of nano-sized domains.

The mesoscale morphology of the material resembles a forest of trees (as seen in figure 1.c) whose characteristic dimensions at different length scales are optimal for light scattering: mean haze in the 400 nm - 700 nm range is up to 85% with a mean transmittance of 90% in the same range, meaning that 85% of the transmitted photons are scattered by the film. We demonstrated how both transparency and haze increase with increasing deposition pressure. However, the electrical conductivity of porous films is extremely poor (higher than 1 M$\Omega$ $\square^{-1}$) due to morphology (i.e., low degree of connectivity) and stoichiometry effects.

## 3.2 Synthesis of functionally graded AZO architectures

We will now describe the synthesis of a functionally graded mesoarchitecture which combines the electrical properties of compact AZO with the effective light management of porous structures. In order to do so, the PLD process starts at an $O_2$ pressure in the 100 Pa - 200 Pa range, which is decreased down to 2 Pa during the deposition time. Since the ablated species impinge on the substrate with higher kinetic energy at lower pressures,[30,31] the architecture was designed in the form of a graded structure with a gradual uniform pressure decrease, in order to prevent damaging of the porous structures by the high-energy species. This gradual transition results in the creation of a "buffer" layer whose growth conditions were optimized as a first, necessary step to obtain functionally graded architectures. An optimal pressure decrease rate of 6.5 Pa/s was found to preserve morphology without excessively increasing film thickness.

The parameters explored in order to obtain the desired combination of properties were the background gas pressure for the deposition of the porous layer and the equivalent thickness (i.e. deposition time) of the porous and compact layers. The deposition pressure of the compact layer was always maintained at 2 Pa, relying on previous works of optimization of TCO functional properties.[26,27] The thickness of the compact layer was chosen in the 250-300 nm range which gives an optimum combination of resistivity (< 1 m$\Omega$ cm) and transparency (> 85%), even though we will show how electrical properties are affected by the morphology of the porous layer. When calculating resistivity from sheet resistance measurements, an equivalent thickness of the compact layer was used, corresponding to the thickness of a layer obtained in an equivalent deposition time. By considering the compact layer as the main source of conductivity we calculated an effective Hall mobility.

**Figure 3** shows cross-sectional SEM images of functionally graded AZO films grown in different conditions. The left part of the figure shows a film whose scattering layer was grown at 160 Pa $O_2$ with an equivalent thickness of 800 nm: the morphology of this layer is characterized by open tree-like structures constituted by hierarchical assemblies of nanosized clusters. This morphology is characterized by a significant fraction of voids, corresponding to a partial spatial separation of the hierarchical structures; as a consequence, the compact layer provides a nonuniform coating which traces out the underlying porous layer surface. In the right part of figure 3, an AZO structure whose bottom layer was grown at 100 Pa $O_2$ is presented; the morphology is significantly less porous and more uniformly connected. An estimate of the mass density of the films taken from deposition rate measurements (with a quartz microbalance) provided a density ratio of about 1/5 between the two bottom layers. The less open morphology results in a more uniform surface, on which the growth of the compact layer occurs in a more uniform manner; the top part is constituted by a flatter, uniform coating.

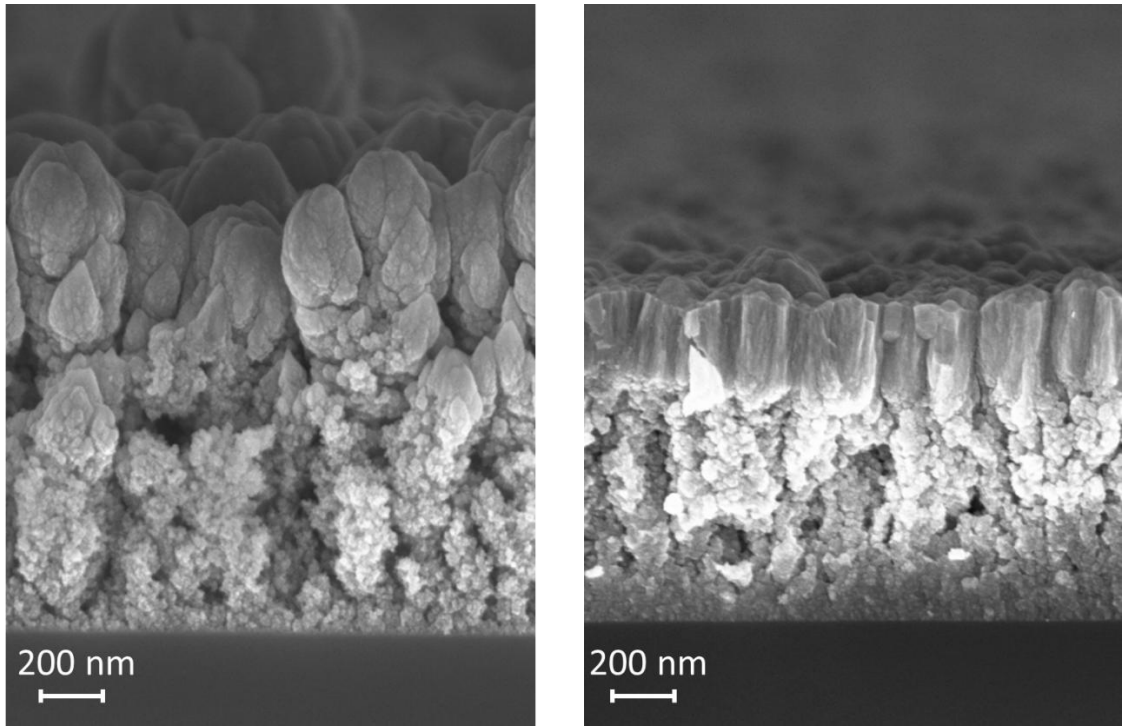

**Figure 3.** Cross-sectional SEM images of graded AZO films with porous layers grown at 160 Pa $O_2$ (left) and 100 Pa $O_2$ (right).

Optical properties were measured in the range from 250 nm to 2000 nm with a UV-Vis-NIR spectrophotometer. An integrating sphere was used for the measurements; for total transmittance the light was detected over the whole solid angle, whereas in the case of diffuse transmittance the specular light was let out of the sphere through an open slit. Haze was calculated as the ratio of diffuse transmittance to total transmittance. Electrical properties were characterized by means of 4-point probe Van der Pauw and Hall effect measurements. An overview of functional properties of the films is reported in the first columns of **Table 1**, which can be used to highlight the main effects of deposition time and pressure.

**Table 1.** Overview of properties of the films shown in Figures 3 and 4. The first two columns refer to Figure 3, the third column displays data obtained from the optimized structure (Figure 4). Mean transmittance and haze have been calculated in the visible (400 nm - 700 nm) range.

| $O_2$ pressure (Pa) | 160/2 | 100/2 | 100/2 (optimized) |
|---|---|---|---|
| Equivalent thickness (nm) | 800+300 | 750+250 | 1600+250 |
| Mean total transmittance | 64.67% | 90.24% | 77.07% |
| Mean haze | 73.08% | 11.61% | 42.39% |
| Sheet resistance ($\Omega \ \square^{-1}$) | 15.1k | 93.1 | 162 |
| Sheet carrier density ($\cdot 10^{15}$ $cm^{-2}$) | 1.68 | 8.2 | 15.2 |
| Hall mobility ($cm^2 \ V^{-1} \ s^{-1}$) | 0.25 | 8.19 | 2.52 |

First, a higher oxygen pressure during the growth of the bottom layer increases its porosity and hence haze; the interaction of the incoming light with the hierarchical structures is maximized when the characteristic dimensions of the "trees" are comparable with the incoming wavelength. An increase in the deposition time of the bottom layer sorts the same effect, since porosity increases along the direction normal to the substrate. On the other hand, a higher effective thickness implies lower total transmittance (in ordered media the decrease would be exponential according to Lambert-Beer's law, in highly scattering systems such as these the optical path is expected to be even longer, with even lower total transmittance). Furthermore, an increase of deposition time or pressure results in higher surface roughness, and hence in a less regular coverage by the compact layer. This has dramatic effects on electrical properties: the sheet resistance of the 160 Pa sample is over 150 times higher than the 100 Pa sample, as reported in table 1. The main reason to this discrepancy is found in the significant decrease in mobility (by a factor 30) due to the lack of a continuous path for charge carriers as a consequence of morphology. Increasing the thickness of the compact layer can significantly increase carrier mobility, but is detrimental for optical transparency: 200 nm of compact AZO (2 Pa) are characterized by the same mean visible transmittance ($\approx$ 90%) as 1 μm of porous AZO (100 Pa).[27,29]

An optimal combination of deposition conditions was identified by increasing the equivalent thickness of a 100 Pa bottom layer to about 1.6 μm, maintaining 250 nm of compact layer on the top. A cross-sectional SEM image of a film grown in these conditions is presented in **figure 4**, and a summary of its functional properties is listed in the last column of Table 1.

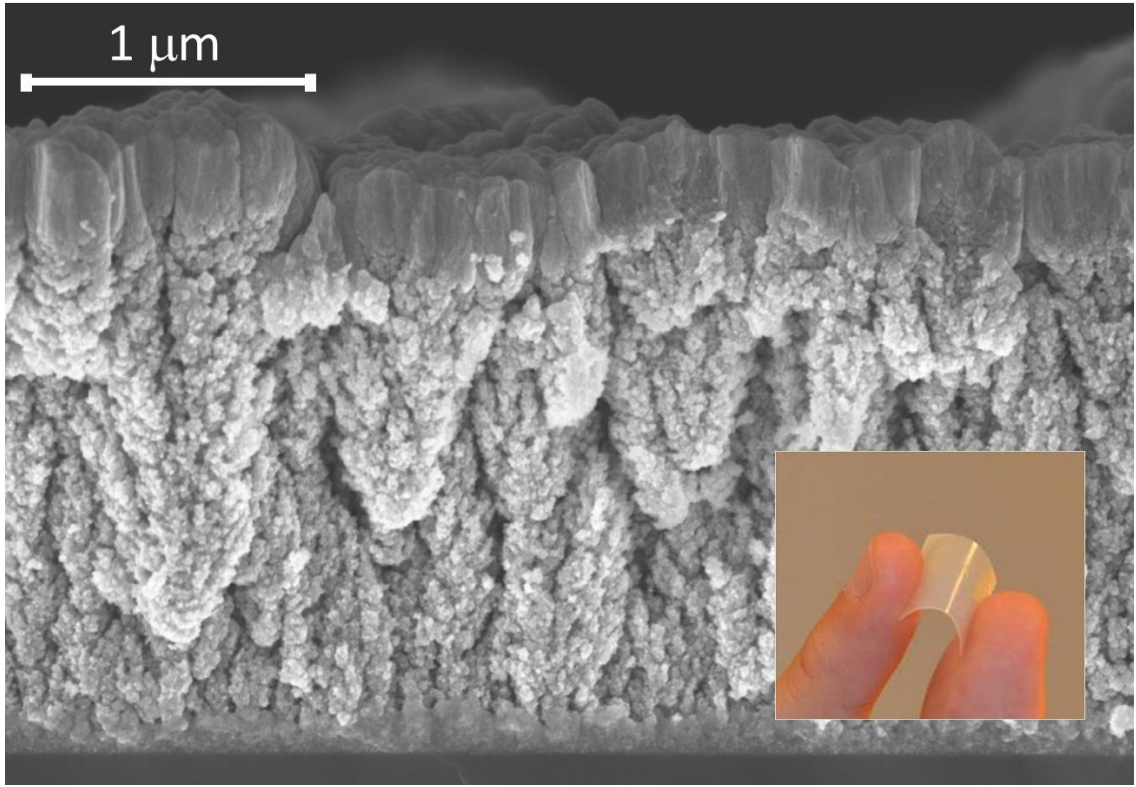

**Figure 4.** Cross-sectional SEM image of an optimized graded AZO structure grown with a thicker bottom layer grown at 100 Pa. The inset on the right shows a photograph of the corresponding sample deposited on a flexible ETFE substrate.

The increase in thickness is sufficient to obtain a haze over 40% without compromising total transmittance, which is close to 80%. We point out that total transmittance can be controlled by choosing the combination of equivalent thicknesses of the two layers. Transmittance and haze spectra are presented in **figure 5**, where total transmittance (dotted line) shows a flat profile in the red-NIR (≈ 80%) and decreases in the blue-UV due to interband electronic transitions, but also in the IR due to free carrier absorption in the conduction band. Light scattering (diffuse transmittance, dashed line) is maximum in the visible (≈ 35%) giving rise to a haze value (dotted line) of the order of 40%. The sheet resistance of this film is around 160 $\Omega$ $\square^{-1}$, with a carrier mobility of around 2.5 cm$^2$V$^{-1}$s$^{-1}$. This is a confirmation of the fact that electrical transport occurs in the compact layer alone: a compact AZO film of the same thickness (200 nm) grown at 2 Pa on a smooth substrate has a sheet resistance which is lower by a factor 5 ($R_S \approx 30\ \Omega\ \square^{-1}$), with a carrier mobility of about 13 cm$^2$V$^{-1}$s$^{-1}$ (5 times higher than the graded film as well). We may thus state that conductivity in graded AZO is morphology-driven, and all the discrepancy between the isolated compact film and the top layer of a graded film is due to nonuniformity of the surface profile (clearly visible from figure 4) which limits carrier mobility. This statement can be supported by Atomic Force Microscopy (AFM) measurements: a linear decrease in carrier mobility with increasing rms roughness was found, with rms roughness values of the order of 40-100 nm. The properties of the films grown on flexible ETFE substrates were also measured: no significant difference was found in optical properties (both transmittance and haze varied by less than 4% with respect to glass substrates) whereas sheet resistance was uniformly increased by a factor 2.5, reaching an optimal value of 400 $\Omega\ \square^{-1}$ in the optimized films.

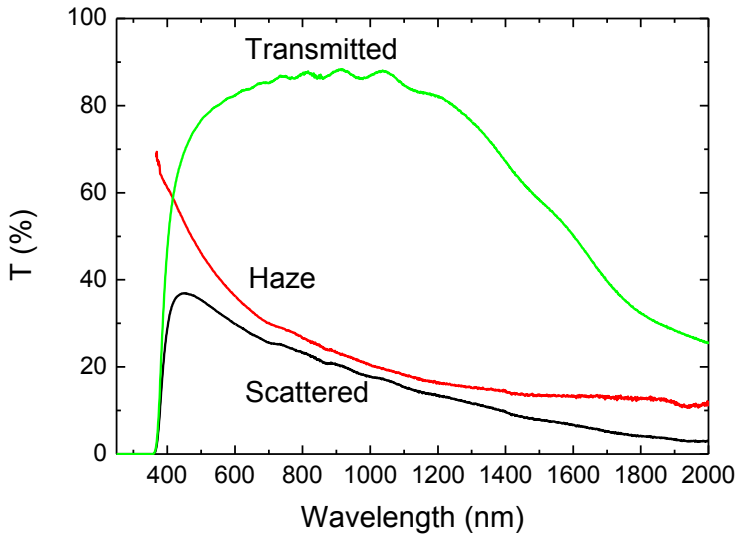

**Figure 5.** Optical transmittance spectra of the optimized graded architecture shown in fig. 4. Total transmittance, diffuse transmittance and haze are reported in the 250 nm - 2000 nm wavelength range. The substrate (glass) contribution to absorption has been corrected for by setting to 1 the light intensity at the glass-film interface.

This increase in resistivity is reasonably due to a decrease in connectivity, as confirmed by SEM analysis (not shown). It is worthwhile to point out that this value does not significantly increase upon bending, probably because of a damping of strain due to the porous structure and buffer layer below the compact film. A photograph of a graded film grown on ETFE is reported in the inset in figure 4.

It is thus possible to design a functionally graded film with the desired combination of electrical and optical properties by controlling the deposition pressure and time of the two layers. The transition rate during the growth of the buffer layer proved to be a critical parameter that must be optimized carefully. We remark that the entire fabrication process is carried out during one single deposition, and no post-treatment is required.

### 3.3 Light harvesting properties

In the framework of advanced solutions for solar cells, semiconducting polymers offer excellent light harvesting capabilities and good charge carrier mobility in the same material. However, in contrast to inorganic absorbers, the energy bands are relatively narrow, and low band gap polymers tend to incompletely absorb light in the visible region of the spectrum. Also, in contrast to inorganic absorbers, a heterojunction is required between the light absorbing polymer and an electron acceptor in order to ionize the photoinduced excitons. For all organic solar cells, "panchromaticity" is achieved by employing an electron acceptor which also absorbs visible light. However, the choice of effective visible absorbing electron acceptors is currently limited to one, namely (6,6)-phenyl-C70-butyric acid methyl ester (C70-PCBM), or derivatives thereof, which is both challenging to isolate and purify and also limited in its own spectral width. The beneficial effects of the scattering layer on the material performance in photovoltaic devices have been tested by evaluating the increase in absorption of Poly[2,1,3-benzothiadiazole-4,7-diyl[4,4-bis(2-ethylhexyl)-4H-cyclopenta[2,1-b:3,4-b']dithiophene-2,6-diyl]] (PCPDTBT), a low bandgap polymer employed in state-of-the-art organic photovoltaics as an electron donor.[32,33] Theoretical studies predict maximum device efficiency if the active medium thickness is below 100 nm.[34] For these reasons, in order to investigate the benefits deriving from light scattering, a 40 nm thick layer of PCPDTBT was spin-coated on a 200 nm thick compact AZO layer and on an optimized 100/2 Pa graded AZO film; optical absorption was measured in the two systems, and the substrate and AZO contributions were subtracted to analyze the effects of haze. The spectra are reported in **figure 6**. The optical density of the polymer deposited on the graded system is significantly increased (up to 100%) in the 400 nm - 600 nm region with respect to the compact one. This corresponds to an increase in light harvesting efficiency from 36% to 60% in the short wavelength absorption peak.[35] In this spectral range the polymer absorption is lower than in the red-NIR regions, and the haze of the graded structure is maximum (see figure 5). It is thus possible to state that the increase in optical density is due to an increase in optical path due to the improved light scattering of the graded structure. This benefit can be expected to increase further if the incidence angle is varied from the normal direction.

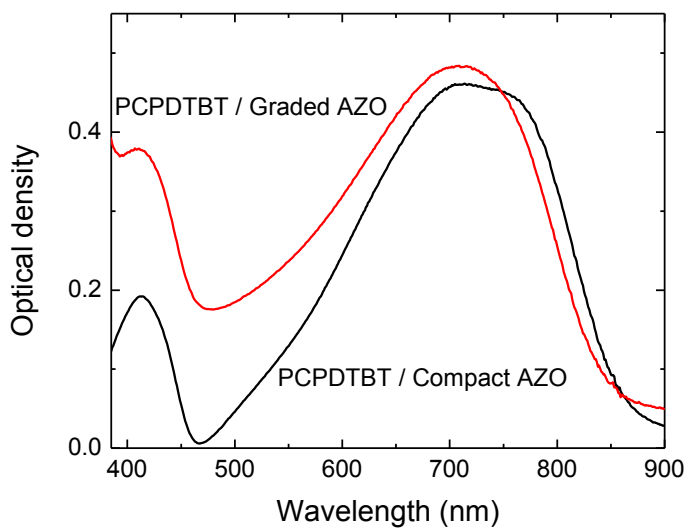

**Figure 6.** Optical density of a PCPDTBT layer on a compact AZO films and on a 100/2 graded AZO architecture. The absorption of substrate and AZO was subtracted from the curves.

## 4. Conclusion

We have demonstrated the synthesis and control of a functionally graded diffusive TCO material, which is simultaneously transparent (80% in the visible range), conducting (160 $\Omega$ $\square^{-1}$ sheet resistance), and efficient in terms of light scattering (40% in the visible). We demonstrated successful adhesion also on flexible substrates with comparable optical properties and moderately lower electrical performance, however we also demonstrated satisfying mechanical stability of the structure on polymer substrates upon bending. The benefits of haze in terms of performance in photovoltaic devices were tested by studying the increase of absorption of a low bandgap polymer due to light scattering alone: in the spectral region where haze is highest, the optical density of the polymer was increased up to 100%. An inversely graded structure (evolving from compact to porous) could also be of potential interest as a light harvesting substrate for inverted bulk heterojunction solar cells.

SUPPLEMENTARY DATA

**Supporting Information** include data relating to thickness optimization, AFM images of the sample surface, SEM image of the sample/polymer interface and transmittance spectra of samples grown on different substrates.

AUTHOR INFORMATION

**Corresponding Author**

*Carlo S. Casari
Dipartimento di Energia and NEMAS – Center for NanoEngineered Materials and Surfaces,
Politecnico di Milano
Via Ponzio 34/3, 20133 Milano, Italy
Phone +39 02 2399 6331
Fax    +39 02 2399 6309
e-mail carlo.casari@polimi.it

ACKNOWLEDGMENT

The authors would like to acknowledge Giulia Grancini for polymer deposition and useful discussion, and Ana Pillado Pérez for AFM measurements. P.G. would like to acknowledge funding from Fondazione Cariplo (PANDORA project).

[35] Light harvesting efficiency is defined as $1-10^{-OD}$, where OD is the optical density